\documentclass[%
preprint,
superscriptaddress,
amsmath,amssymb,
aps,
prl,
]{revtex4-2}
\usepackage{color}
\usepackage{gensymb}
\usepackage{graphicx}
\usepackage{dcolumn}
\usepackage{bm}
\usepackage{hyperref}
\usepackage{epstopdf}
\usepackage{float}
\begin{document}


\title{Stringiness of Hyaluronic Acid Emulsions}

\author{H. V. M. Kibbelaar}
 \affiliation{Van der Waals-Zeeman Institute, Institute of Physics, University of Amsterdam, 1098 XH Amsterdam, The Netherlands.}
\author{A. Deblais}
 \affiliation{Van der Waals-Zeeman Institute, Institute of Physics, University of Amsterdam, 1098 XH Amsterdam, The Netherlands.}
 \affiliation{Unilever Innovation Centre Wageningen, Bronland 14, 6708 WH Wageningen, The Netherlands.}
\author{K. P. Velikov}
\affiliation{Van der Waals-Zeeman Institute, Institute of Physics, University of Amsterdam, 1098 XH Amsterdam, The Netherlands.}
\affiliation{Unilever Innovation Centre Wageningen, Bronland 14, 6708 WH Wageningen, The Netherlands.}
\author{D. Bonn}
\affiliation{Van der Waals-Zeeman Institute, Institute of Physics, University of Amsterdam, 1098 XH Amsterdam, The Netherlands.}
\author{N. Shahidzadeh}
\affiliation{Van der Waals-Zeeman Institute, Institute of Physics, University of Amsterdam, 1098 XH Amsterdam, The Netherlands.}

\date{\today}

\pacs{Valid PACS appear here}
\maketitle

\section{Abstract}

\textbf{Objective}
Cosmetic emulsions containing hyaluronic acid are ubiquitous in the cosmetic industry. However, the addition of (different molecular weight) hyaluronic acid can affect the filament stretching properties of concentrated emulsions. This property is often related to the ``stringiness'' of an emulsion, which can affect the consumer's choice for a product. It is thus very important to investigate and predict the effect of hyaluronic acid on the filament stretching properties of cosmetic emulsions.

\textbf{Methods}
Model emulsions and emulsions with low and high molecular weights are prepared and their filament stretching properties are studied by the use of an extensional rheometer. Two different stretching speeds are employed during the stretching of the emulsions, a low speed at 10 \textmu m/s and a high speed at 10 mm/s. The shear rheology of the samples is measured by rotational rheology. 

\textbf{Results}
We find that filament formation only occurs at high stretching speeds when the emulsion contains high molecular weight hyaluronic acid. The formation of this filament, which happens at intermediate states of the break-up, coincides with an exponential decay in the break-up dynamics. The beginning and end of the break-up of high molecular weight hyaluronic acid emulsions show a power-law behaviour, where the exponent depends on the initial stretching rate. At a lower stretching speed no filament is observed for both high molecular weight and low molecular weight hyaluronic acid emulsions, and the model emulsion. The emulsions show a power-law behavior over the whole break-up range, where the exponent also depends on the stretching rate. No significant difference is observed between the shear flow properties of the emulsions containing different molecular weights hyaluronic acid. 

\textbf{Conclusion}
In this work, we underline the importance of the molecular weight of hyaluronic acid on the elongational properties of concentrated emulsions. The filament formation properties, e.g. the stringiness, of an emulsion is a key determinant of a product liking and repeat purchase. Here, we find that high molecular weight hyaluronic acid and a high stretching speed are the control parameters affecting the filament formation of an emulsion.

\textbf{Keywords}
Emulsions,
Hyaluronic acid,
Extensional rheology,
Shear rheology,
Stringiness,
Filament formation,
Complex emulsions.

\textbf{Acknowledgements}
This research was performed within the framework of the ``Molecular aspects of biopolymers defining food texture perception and oral digestion'' project funded by NWO (The Netherlands Organization for Scientific Research), grant number 731.017.201, Unilever and Anton Paar. The project partners (Wageningen University \& Research, Unilever, Anton Paar) have contributed to the project through regular discussion.


\newpage

\section{Introduction}

Hyaluronic Acid (HA) has become one of the most used ingredients in cosmetic products. It is widely used as a moisturizing active ingredient in cosmetic formulations to enhance skin hydration and elasticity, resulting in softer, smoother and radiant skin \cite{son2017hyaluronan}. HA is a natural occurring linear polysaccharide composed of repeating disaccharide units of D-glucuronic acid and N-acetyl-D-glucosamine \cite{fraser1997hyaluronan} and it is one of the main components of the extra cellular matrix in which cells and fibrous constituents of the matrix such as elastin and collagen are embedded \cite{kielty1992type,baccarani1990immunocytochemical}. HA has many biological functions including maintenance and viscoelasticity of liquid connective tissues such as synovial joints and the eye vitreous, contribution in tissue regeneration where it assures normal tissue function and supporting the regulation of wound healing \cite{laurent1995functions,kogan2007hyaluronic,lapcik1998}. Another unique character of HA is its extremely high water binding capacity which ensures good tissue hydration and water transport in the human body. For example, HA within skin layers is able to hold up 1000 times its weight in water molecules \cite{cleland1970ionic}. These extraordinary properties make HA a useful ingredient for cosmetic applications. 

HA is available in many different molecular weights (ranging from ten thousands to millions of Daltons) and the molecular weight of HA needs to be considered when formulating a cosmetic product. Where high molecular weight (HMW) HA works as a film forming polymer and thereby contributes to the moisture content of the skin, and the possible decrease of the transepidermal water loss, Low molecular weight (LMW) HA is primarily utilized to improve skin penetration to restore a sustained physiological and hydrated micro environment for optimize skin rejuvenation and tissue repair \cite{smejkalova2015hyaluronan, pavicic2011efficacy, bukhari2018hyaluronic, farwick2011, brown1999absorption}. Thus, the larger the molecular weight of HA, the more dominant the physical chemical properties of the vehicle (e.g. emulsion) are affected. The molecular weight drastically affects the rheological properties of the cosmetic product. The increase in molecular weight reinforces the formation of a three-dimensional network of HA resulting in an increase of the viscosity and viscoelasticity \cite{fallacara2018hyaluronic}. Rheological properties are a crucial parameter in the field of cosmetics, where the consumer's choice for a product highly depends on the texture of the product \cite{wortel2000skin, wang1999effect}. Many various texturing properties characterize cosmetic products \cite{tadros2004application, barnes1994rheology}. Slipperiness, spreadibility, stickiness, thickness are examples of sensory attributes which are used for the characterization and perception of cosmetic products \cite{lee2005terminology, arrieta2020design, savary2019instrumental}. It is usually attempted to link a  rheological parameter (e.g. shear viscosity, storage and loss moduli) to such sensory perceptions determined by a trained test panel \cite{meilgaard2006sensory, ahuja2019rheological, ozkan2020rheological, estanqueiro2016comparison, gilbert2013predicting}. 

The sensory attribute correlated to the extensional properties of a cosmetic product is often expressed as the  ``cohesiveness" or ``stringiness" and is assessed by the consumer during the pick-up, which corresponds to the properties perceived in the hand when the product is taken from the container or massaged between the finger tips. The stringiness of a product is defined by Civille and Dus \cite{civille1991evaluating} as  ``amount of sample deforms or strings rather than breaks when fingers are separated", i.e. the stringier a product, the more prone the filament formation. Filament formation of fluids is a property observed and measured during extensional rheology \cite{mckinley2005visco, deblais2018pearling, amar2005fingering}. Among the various texturing properties characterizing cosmetic products, little is known about the cohesiveness/stringiness attribute for concentrated emulsions containing different molecular weights of HA. HA solutions are known to show a strong viscoelastic response with the formation of filaments in the elongational direction \cite{kibbelaar2020capillary, haward2013extensional}. However, for its aforementioned applications in cosmetic products, HA is often added to other products like emulsions \cite{olejnik2015stability}. Earlier work of Louvet \textit{et al.} \cite{Louvet2014} investigates the elongational dynamics of a simple model emulsion. However, the addition of a polymer to an emulsion and its effect on the break-up dynamics remains unknown. In this work, we examine the effect of different molecular weight HA on the rheological properties of a concentrated model emulsion, where we focus on the elongational properties by using a filament stretching rheometer. We also investigate the effect of stretching speed as this is relevant during the use of the emulsion by the consumer. 

\section{Materials and Methods}

\textbf{Extensional rheology.} To study the extensional thinning of the emulsions we use a custom built filament stretching rheometer (Anton paar, MCR 301), similar to the one described in \cite{Huisman2012, Louvet2014}. A small sample of 40~\textmu L fluid is initially placed between two circular end plates ($D_{0}$ = 5~mm and $L_{0}$ = 2.5~mm) which are moved apart at a constant velocity until the bridge breaks. In this work we employ two speeds to study the effect of the stretching speed on the dynamics of the emulsions: 10~\textmu m/s (low speed) and 10~mm/s (high speed). The evolution of the liquid bridge is recorded with a fast camera (Phantom V7) allowing frame rates up to 10 000 frames per second. The camera is equipped with a microscope tube lens, with an objective up to 12x magnification (Navitar) that allows a spatial resolution of 3~\textmu m per pixel. The whole setup is placed in a closed chamber and continuously flushed with humid air (80$\%$ RH) to prevent evaporation during the measurements. 

\textbf{Shear rheology.} Shear rheology measurements were performed with a stress-controlled rheometer (Anton Paar MCR 302), equipped with a roughened cone plate geometry with a diameter of 50~mm and cone angle of 1$^{\degree}$. The experiments were performed at a temperature of 22~$^{\degree}$C set by a Peltier system. A humidity chamber around the geometry allow us to suppress evaporation during the whole measurement. The steady shear experiments were performed by carrying out a shear rate sweep from 1 $\cdot$ 10$^{-3}$ to 1 $\cdot$ 10$^{2}$ s$^{-1}$. Stresses were averaged over 15 s after reaching the steady state. 

\textbf{Emulsion preparation.} The emulsions (with and without HA) are prepared by diluting a stock castor oil in water/SDS emulsion of droplets with a total internal volume fraction of $\phi_{oil}$ = 0.90 using 1 wt. $\%$ aqueous SDS solutions until a final volume fraction of $\phi_{0il}$ = 0.80 castor oil droplets is reached. The stock emulsion is prepared by addition of the castor oil to the aqueous phase and stirred with a Silverson L5M-A emulsifier at 2,500 rpm for 6 minutes, followed by stirring at 6000 rpm for 3 minutes and 10.000 rpm for 2 minutes. Both castor oil and SDS are purchased from Sigma Aldrich. The emulsions with HA are then made by diluting the stock emulsion, immediately after preparing, with an aqueous solution containing 2 wt. $\%$ SDS and 1 wt. $\%$ HA, so that the final emulsions have a volume fraction of $\phi_{0il}$ = 0.80, and contain 0.5 wt $\%$ HA (an HA concentration typically used in cosmetic formulations) and 1 wt. $\%$ SDS. The final oil volume fraction of all studied emulsions in this work is  $\phi_{0il}$ = 0.80 and are stabilized by 1 wt. $\%$ SDS. This protocol of diluting a stock emulsion is chosen so that breakage of HA chains during the emulsion preparation process is avoided. This procedure also allows to prepare HA-containing emulsions by diluting the stock emulsion with only 1 wt. $\%$ SDS. The aqueous HA/SDS solutions for the dilutions were prepared following a frequently used protocol \cite{Burla, Giubertoni2019, kibbelaar2020capillary}: 1 wt. $\%$ HA solutions were prepared using hyaluronic acid sodium salt powder from Streptococcus equi bacteria (HMW: 1.5–1.8 MDa, Sigma Aldrich and LMW: 21-40 KDa, Lifecore) in distilled water, together with a fixed NaCl concentration (0.15 M). Samples were vortexed for a few seconds to ensure mixing of the ingredients and homogenized under modest rotation during a period of 5 days. 1 wt. $\%$ of SDS was added after these 5 days and stirred for 1 day. 

\textbf{Confocal microscopy.} The droplet size of the emulsions is studied by confocal  fluorescence microscopy (Leica TCS SP8), using fluorescent molecules rhodamine (Sigma-Aldrich) in the continuous water phase. Rhodamine is excited at 500 nm and emits at 540 nm.

\section{Results and discussion}
\begin{figure}[h]
\includegraphics[scale=1]{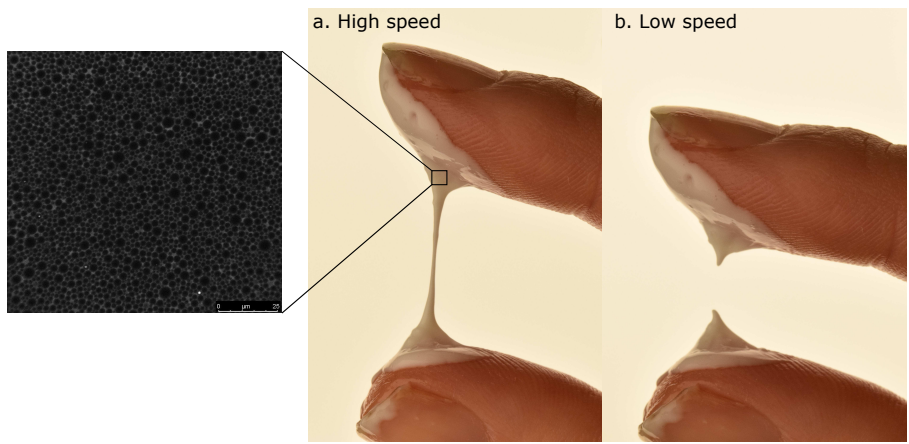}
\caption{Stretching of the HMW HA emulsion between the fingers at (a) low speed and (b) high speed. From this visual observation can be seen that the degree of ``stringiness" depends on how fast the consumer moves the fingers while using the product. The inset shows a typical fluorescence confocal microscope image of the HMW HA emulsion.}
\label{F1}
\end{figure}

Figure \ref{F1} shows the stretching of a model emulsion containing 0.5 wt. $\%$ HMW HA, between two fingers, at two different speeds: (a) low (b) high. Where low and high are considered as stretching slower and faster by the opinion of a consumer. The inset of the figure shows a typical confocal microscope image of the oil droplets dispersed in the continuous phase of HA and 1 wt. $\%$ SDS. The oil droplets have a mean diameter of 2.5 $\pm$ 0.6 \textmu m, as measured with confocal microscopy. 1 wt. $\%$ SDS stabilized castor oil-in-water emulsions are chosen as the model system as this system is commonly used to study the rheology of model suspensions. Studies have shown that  1 wt. $\%$ SDS stabilizes concentrated ($\phi_{oil}$ = 0.80) castor oil-in-water emulsions \cite{becu2006yielding, Louvet2014, huang2008shear, deblais2015spreading, paredes2013rheology}. The figure clearly shows the stringiness of the emulsion, which is often observed during the use of a product by the consumer: at high speeds a long filament of emulsion is formed, while at low speeds no filament is present. In order to evaluate this behaviour more precisely, we study both shear and extensional rheology. The shear flow of cosmetic emulsions is one of the common measured flows when assessing a cosmetic product. Figure 2 shows the shear rheology of the standard emulsion, the LMW HA emulsion and the HMW emulsion. All three samples show shear thinning behaviour, where no significant difference in viscosity is observed for the range of shear rates investigated for the HMW emulsion: The effect of the addition of HA polymers on the stringiness (Fig. \ref{F1}) of the emulsion cannot be explained from the shear flow measurements.  

\begin{figure}
\includegraphics[scale=1]{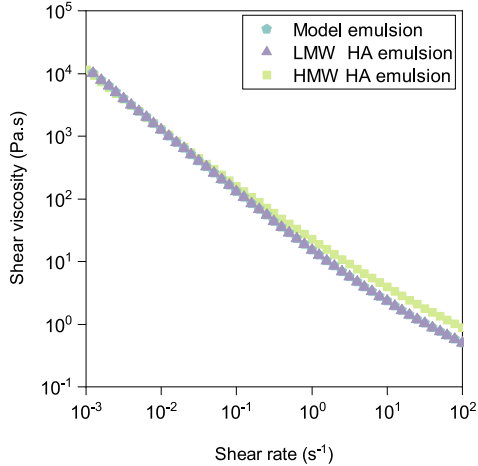}
\caption{Shear viscosity as a function of shear rate of the model emulsion and the emulsion with addition of different molecular weights HA. No significant difference is observed between the shear viscosity of the two different samples.}
\label{F4}
\end{figure}

We thus further study the stringiness with the help of a filament stretching rheometer. We first focus on the emulsions containing HMW, as HMW HA is more likely to effect the elongational properties \cite{kibbelaar2020capillary}. Figure 3(a) shows typical photographs of the break-up dynamics for the emulsion with HMW HA at low (10~\textmu m/s) and high speeds (10~mm/s), and the corresponding thread radius as a function of the rescaled time is plotted in Fig. 3(b). Thinning dynamics for so-called power-law shear thinning fluids, including yield stress fluids which follow the Herschel-Bulkley equation \cite{paredes2013rheology}, are universal and should follow  $h_{min} \propto \tau^n$ \cite{Huisman2012,yarin2004elongational}. Recent experiments support this thinning law by finding thinning dynamics to be well described by the exponent $n$ \cite{Huisman2012, yarin2004elongational, aytouna2013drop}, where the exponent is also stretching rate dependent \cite{Louvet2014}. 

\begin{figure*}
\includegraphics[scale=1]{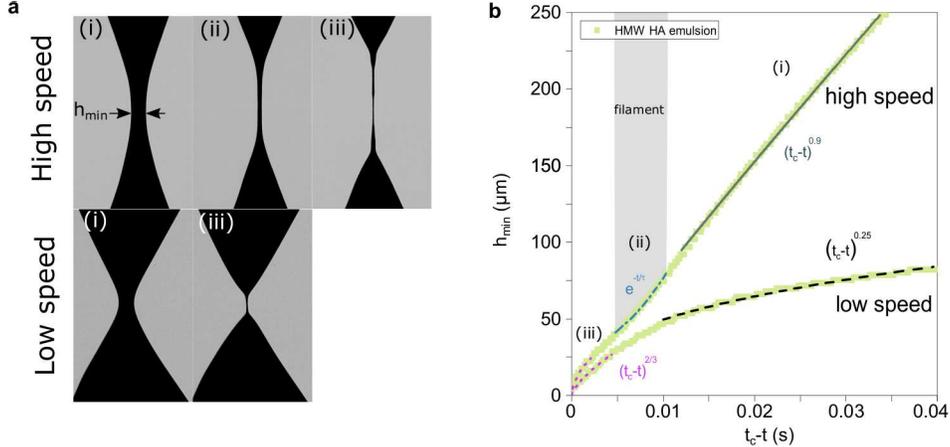}
\caption{(a) Photographs of the neck break-up dynamics of HMW HA emulsions at a high speed (10~mm/s) and low speed (10~\textmu m/s). (b) Minimum filament diameter ($h_{min}$) as a function of the rescaled time $t_{c}-t$. Numbers correspond to the photographs in panel (a). The red lines are fits with power law of 2/3 just before the break-up, and at longer timescale fits with a power-law of 0.25 and 0.9 for respectively low speed and high speed. The orange line at intermediate time is an exponential fit, indicating the formation of a filament.}
\label{F2}
\end{figure*}

In our experiments we distinguish two regimes for the HMW HA stretched at low speeds. The curve can be fitted with a power-law over the whole range, however the value of the exponent $n$ changes at the last instants before break-up from a value of 0.25 (i) to a value of 2/3 (ii). A 2/3 power law just before break-up is also observed for other suspension/dispersion systems \cite{harich2016depletion, pan2015drop, miskin2012droplet}. In these systems a local decrease in  suspension/dispersion volume fraction is generally observed, causing the solvent properties to determine the dynamics \cite{zhao2015inhomogeneity, mathues2015capillary}. At high stretching speeds, the dynamics can be divided into three regimes. At the later times (i), the break-up dynamics can be well described by $t_{c}-t^{0.9}$, where the exponent ($n$ = 0.9) has a higher value than for the low stretching speed, in agreement with the findings of Louvet \textit{et al.} \cite{Louvet2014}. There is no quantitative understanding why this happens to date, but a proposed hypothesis is that the difference in fluidity of the emulsion close to the plates and in the fluid neck leads to a nontrivial neck structure \cite{Louvet2014}. Then, at intermediate times (ii) the formation of a filament is observed in the movies of the break-up dynamics. The neck elongates to form a long filament reminiscent of visco-elasto-capillary thinning often observed for polymer solutions \cite{wagner2005droplet}. The time at which the filament starts to form exactly coincides with the occurrence of an exponential decay in the break-up dynamics, and the corresponding break-up dynamics can be fitted with an exponential function. The dynamics of this part can be approximated by $h_{min} \propto A e^{-t/\tau}$, with $\tau$ $\approx$ 8 ms. Then just before break-up (iii) a 2/3 power-law behaviour is also observed. From the capillary break-up experiments can thus be observed that the addition of HMW HA and high stretching speed introduces an extra regime where a visco-elasto-capillary thinning behaviour is induced. This induced exponential thinning behaviour is earlier observed in the work of Harich \textit{et al.} \cite{harich2016depletion}, where the introduction of attractive forces between otherwise non-interacting particles induced similar exponential thinning behavior.

\begin{figure*}
\includegraphics[scale=1]{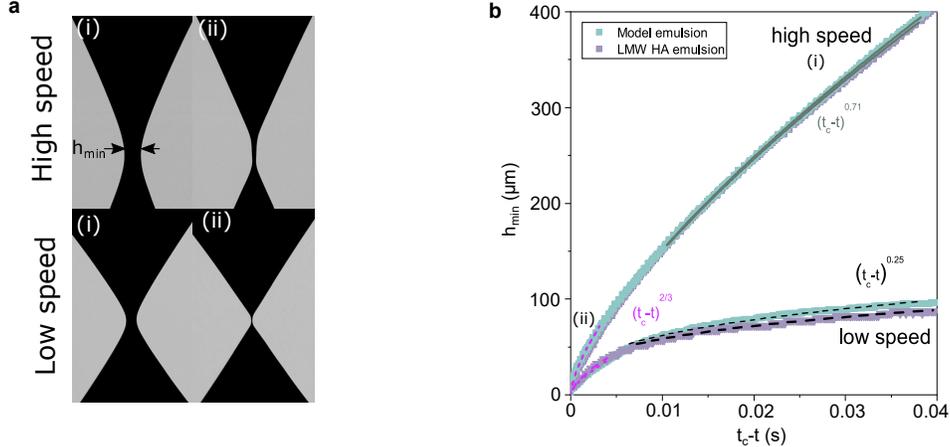}
\caption{Photographs of the neck break-up dynamics of the LMW HA emulsion at a high speed (10 mm/s) and low speed (10 \textmu m/s) (the break-up photographs of the model emulsion are similar to the LMW HA emulsion). (b) Minimum filament diameter ($h_{min}$) as a function of $t_c-t$ (rescaled time to break-up), also of the model emulsion. Numbers correspond to the photographs in panel (a). The red lines are fits with a power law of 2/3 just before the break-up, and at longer timescale fits with a power-law of 0.25 and 0.71 for respectively low speed and high speeds. No exponential function can be fitted, which is in agreement with the absence of a filament.}
\label{F3}
\end{figure*}

In order to better quantify and understand the effect of the molecular weight of HA on the model emulsion, the emulsions with LMW and the model emulsion without added HA polymer are studied in the same way as the HMW sample. Figure~4(a) shows typical photographs of the break-up dynamics for the emulsion with LMW HA at low and high speeds, and the corresponding thread radius as function of time is plotted in Fig.~4(b). This plot also contains the break-up dynamics of the standard emulsion stretched at low and high speed; both emulsions follow the same trend (the photos of the break-up dynamics of the model emulsion are not shown as they are similar to the ones of the LMW HA emulsion). For both (low and high speeds) curves of the LMW and model emulsion two regimes are observed, where a power-law regime is observed over the whole range, in contrast to the HMW HA emulsions. At low speeds, the dynamics follow the same power-law behaviour as the HMW emulsions stretched at low speed, with similar values of the exponent \textit{n} at the beginning (i) and end (ii) of the break-up (0.25 and 2/3, respectively). The LMW HA emulsion stretched at a high speed is also well fitted with a power-law, where for regime (i) an exponent of 0.71 is found, and for the part just before break-up (ii) a 2/3 is found. The exponent 0.71 is slightly different from the exponent found for the HMW in the same regime due to the difference in molecular weight HA of the two samples. From the photos of the break-up dynamics no filament formation is observed and an exponential regime is also not present in the thinning dynamics. This demonstrates that the LMW HA solutions do not exhibit a pronounced visco-elasto-cappilary thinning behaviour, regardless of the speed.

\section{Conclusion}

In conclusion, we studied the effect of different molecular weight HA on the elongational properties of emulsions to get more insight into the effect of HA on the filament stretching properties (stringiness) of an emulsion. We found that filament formation only occurs when HMW HA emulsions are stretched at high speeds, where the time of filament formation coincides with an exponential behaviour in the thinning dynamics. Both HMW, LMW and the model emulsion show power-law behaviour at the beginning and end of the break-up where the exponent in the first regime depends on the initial stretching rate. All three emulsions, regardless of the stretching speed, show a 2/3 power law behaviour just before break-up. The effect of molecular weight on the shear flow of the model emulsions, is not as significant. This work shows that while formulating cosmetic emulsions with different molecular weight HA, the elongational flow (and filament formation) as a function of molecular weight is more important for the stringiness of the emulsion than the shear thinning behaviour.

\section{Acknowledgements}

A.D. acknowledges the funding from the Horizon 2020 program under the Individual Marie Skłodowska-Curie fellowship number 798455.

\end{document}